\RequirePackage{ifpdf}

\documentclass{pCHEF}
\title{Experimental Challenges of the European Strategy for Particle Physics}

\ShortTitle{HL-LHC Challenges}

\author{\speaker{Sebastian White}\thanks{ representing C. Lu and K. McDonald(Princeton), T. Tsang(Instr. Div), M.McClish, R. Farrell(RMD/Dynasil).}\\
        Center for Studies in Physics and Biology, The Rockefeller University, NY, USA\\
        E-mail: \email{swhite@rockefeller.edu}}

\abstract{In planning for the Phase II upgrades of CMS and ATLAS major considerations are: 1)being able to deal with degradation of tracking and calorimetry up to the radiation doses to be expected with an integrated luminosity of 3000 $fb^{-1}$ and 2)maintaining physics performance at a pileup level of ~140. Here I report on work started within the context of the CMS Forward Calorimetry Task Force and continuing in an expanded CERN RD52 R$\&$D program integrating timing (i.e. measuring the time-of-arrival of physics objects) as a potential tool for pileup mitigation and ideas for Forward Calorimetry. For the past 4 years our group has focused on precision timing at the level of 10-20 picoseconds in an environment with rates of $~10^6-10^7$ Hz/$cm^2 $
as is appropriate for the future running of the LHC (HL-LHC era).  A time resolution of 10-20 picoseconds is one of the few clear criteria for pileup mitigation at the LHC, since the interaction time of a bunch crossing has an rms of 170 picosec. While work on charged particle timing in other contexts (i.e. ALICE R$\&$D) is starting to approach this precision, there have been essentially no technologies that can sustain performance at these rates. I will present results on a tracker we developed within the DOE Advanced Detector R$\&$D program which is now meeting these requirements. I will also review some results from Calorimeter Projects developed within our group (PHENIX EMCAL and ATLAS ZDC) which achieved calorimeter timing precision< 100 picoseconds.}

\FullConference{Calorimetry for High Energy Frontiers - CHEF 2013,\\
		April 22-25, 2013\\
		Paris, France}

\usepackage{CHEF2013_definitions}
\begin{document}

\section{Strategy}

The year 2012 was a very good year at the LHC. During 2012 the LHC achieved ~80$\%$ of design luminosity, the 
experiments recorded a useful integrated luminosity of $\sim22$ $fb^{-1}$ and, of course, significant analyses were completed.
For those of us fortunate enough to have been at CERN during this period it was quite apparent that the community was running
flat out in terms of human resources, computing power and complexity of the data (due to pileup). 

	Since the European Strategy for Particle Physics calls for a 10-fold increase in this level of performance at the LHC (i.e. a 10-fold 
higher integrated luminosity than expected from a 15 year extrapolation of 2012 performance) it is worth reviewing how the 
community could address this challenge.

\section{Challenges}

	The primary challenge is the increase in intensity implied by the 3000 $fb^{-1}$ target. There has been a long debate in High Energy Physics about the
limits of viability of general purpose experiments at high intensity. Particularly interesting is the paper by Huson, Lederman and Schwitters \cite{Huson}
%%give citation in Snowmass Proceedings. This year (2013) marks the 30th anniversary of the termination of Isabelle and the 20th anniversary of the
%% termination of the SSC- interresting milestones as we discuss the strategy going forward.
written a year before the US community decided to abandon Isabelle(whose viability depended on being able to push this intensity limit) in favor of a higher energy machine (SSC).

	Over the past 20 years there has been a significant R$\&$D program- both in Europe and the US- to address radiation damage due to the large \underline{integrated}
dose projected at the LHC. As a result, there are now tools in place which allow us to project degradation in performance- mostly due to non-ionizing energy loss ("NIEL")
due to this intense activity.( For example, CMS scaling rules for damage to Avalanche Photo-Diodes (APDs) are very useful for the work described below.)
	
	There has been far less emphasis on the consequences of the high \underline{instantaneous} rates. The more obvious, trigger, challenge is currently getting attention in ATLAS and CMS. 
However there has been very little support for developing tools to mitigate pielup- the fact that a single "frame" recorded in ATLAS or CMS typically includes 30-40 other pp interactions because of
the bunch structure of the beams. Over the next 10 years  the current strategy for LHC will result in a significant increase in pileup- to a level of $\sim$ 140. 

	Event pileup has significant impact on the level of background to be expected in a given analysis of the data- depending on the extent to which the analysis brings together information from
different aspects of the same event. For example, we expect analyses of Higgs $\rightarrow$ 4 leptons to be very robust against pileup. On the other hand,  in tagged "vector boson fusion"- wherein (often
forward) proton remnants in the form of jets are measured in addition to the Higgs decay candidate- there is a high likelihood of mis-associating jets from any of the other 140 events recorded in the same frame with the Higgs candidate. 

	A striking example of this is the current effort by the TOTEM experiment to analyze inelastic diffractive data in association with information from the central rapidity region (recorded in the CMS data stream). They make a data selection requiring conditions with <$3\%$ pileup and, even under these conditions, find a significant level of pileup induced background. Since it is perhaps reasonable to hope for a 10-fold reduction of the effects of pileup using event timing (see discussion below) members of TOTEM are planning on possible future running with a luminosity about an order of magnitude lower in point 5 than was achieved in 2012.
	
	For the past 5 years our group has been involved in developing tools to mitigate pileup, partially under a DOE advanced detector R\& D grant \cite{doeadrd}. Surprisingly, this is perhaps the only US supported R$\&$D on this critical generic issue.
	
\section{Pileup in Space and Time}
	To understand which tools might be brought to bear on the problem of pileup mitigation it is useful to simulate the interaction of 2 LHC bunches as they collide in one of the 2 high luminosity regions- points 1$\&$ 5 (ATLAS and CMS). These 2 regions have the same design except that the beam crossing angles -nominally 150 micro-radians- are oriented in the vertical and horizontal planes, respectively. Such a study was performed in 2007\cite{correlation} using the original LHC design book parameters.
	
	\begin{figure}[h!] 
\begin{center}
\includegraphics[width=1\textwidth]{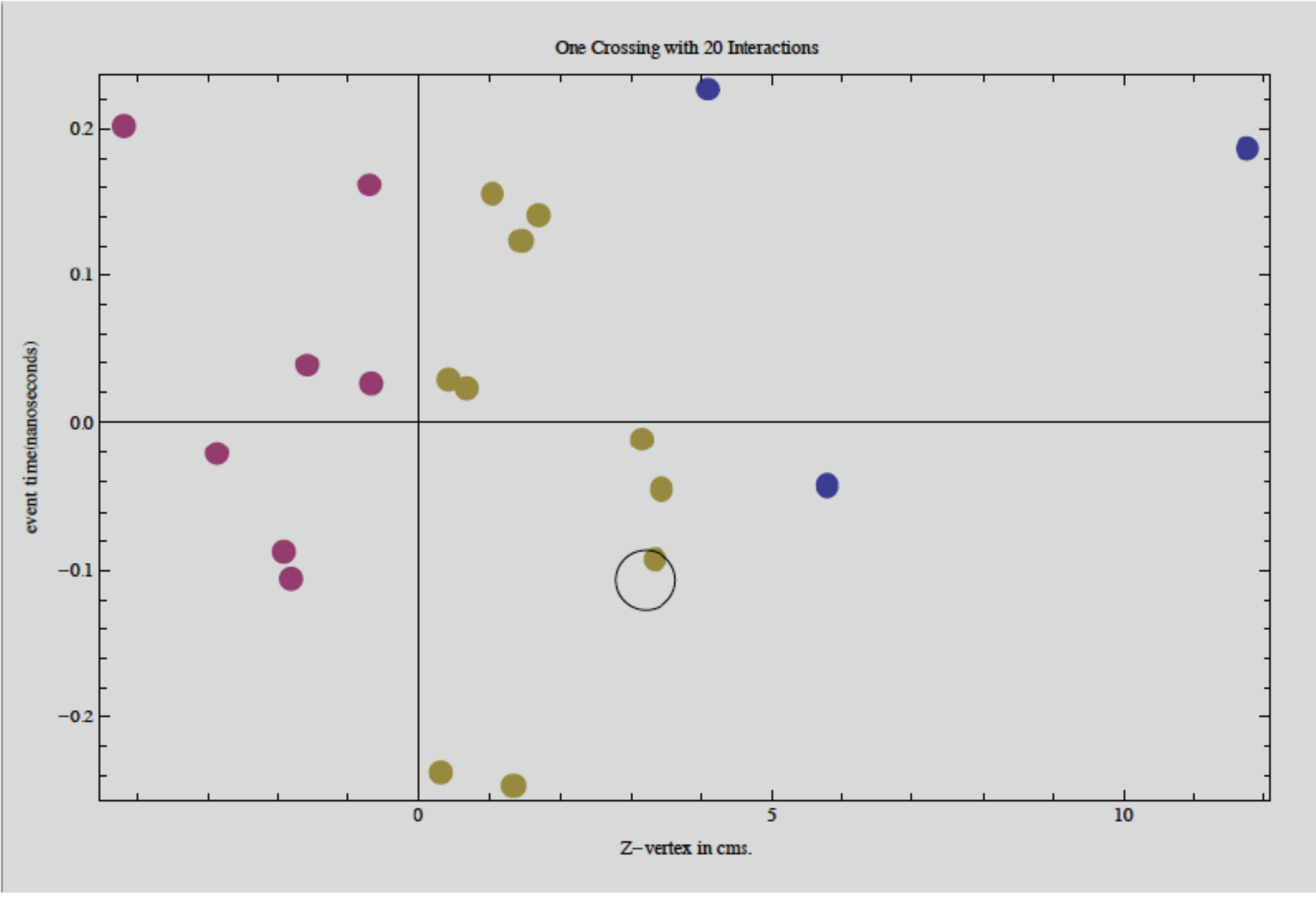}
\caption{20 Events occurring within the same bunch crossing and their coordinates in time-of-occurence and z-vertex location. The open circle represents a possible error ellipse on the association with forward fragments from one of these events} 
\label{fig1} 
\end{center}
\end{figure}

	Individual pp collisions can be visualized as occurring in the 2 dimensional space of z-vertex location and time-of-occurence within an overall envelope of the bunch traversal, which has an rms of 170 picoseconds (=$\sigma_L^t$)-"event time". A typical bunch crossing is shown in Fig.~1, where a pileup of 20 was assumed, in the space of event time and z-vertex. In this case 2 forward remnants were identified and it is assumed that the time of arrival of each remnant (one in the forward and one in the backward hemisphere) can be measured to a precision of $\sim\frac{\sigma_L^t}{10}$. Since the time-of-arrival difference between these remnants is related to the z-vertex location ($\delta$t=2z/c in the case of $0^o$)  this would allow association with the corresponding vertex for activity in the central region- that of a Higgs decay candidate, for example. Similarly the time average of the fragments-giving the time of event occurrence could be compared to that of a central object.
	
	In addition to the actual event times and positions, denoted by filled circles, an error ellipse indicates the possible resolution in these parameters from fragments in the opposite hemispheres (assuming the above jet reconstruction resolution). In this particular event it can be seen that event time in the central region (i.e. from 2-$\gamma$ or 4 lepton reconstruction of a Higgs decay) would help
to resolve ambiguities in, for example, a VBF Higgs production candidate.

	From this simulation exercise one recovers the following non-intuitive principles which must hold in the design book case:
	\begin{itemize}
\item{due to the Poisson distributed density distribution, given a candidate event in position and time, the next-nearest candidate's distance has probability distribution which is exponentially falling in distance and time from zero. This can also be derived from queuing theory\cite{correlation}.}
\item{The Luminosity distribution in space and time factorize- i.e. the space distribution of the luminosity is invariant in time.
	
\begin{equation}	
L(z,t)=I(z,t)\cdot I(z,-t)=\frac{e^{-\frac{(-ct+z)^2}{2\sigma_l^2}-\frac{(ct+z)^2}{2\sigma_l^2}}}{2\pi\sigma_l^2}
	=\frac{e^{-\frac{-c^2t^2+z^2}{\sigma_l^2}}}{2\pi\sigma_l^2}
	 =L(z)*L(t)
	\end{equation}	
}
	\end{itemize}
	
	The same simulation can then be applied to new concepts, partly based on "crab cavities",  which could be used to alleviate pileup at the level of  $\sim$a factor of 2\cite{Fartoukh}.
	
	Measurement of the z-vertex is generally of high quality in ATLAS and CMS, due to high resolution silicon trackers, but becomes essentially impossible in the very forward direction, beyond the nominal tracking region ($\eta>2.5$). Similarly for $\gamma$'s in CMS, which has not been optimized for $\gamma$ directional measurement, vertex resolution is limited.
	
	The time resolution needed for pileup mitigation is clearly $\sigma_t\sim\frac{\sigma_L^t}{10}\sim20$ picoseconds, since it must be about an order of magnitude smaller than the luminosity envelope in time to be effective.
	
\section{Timing for Pileup Mitigation}

	Current use of the ATLAS and CMS detectors does not emphasize timing beyond association of measurements with the correct bunch crossing. Nevertheless there have been relevant studies of time resolution for jets in ATLAS and CMS with existing data (see H. de la Torre, these proceedings, wherein an ATLAS LAr $\sim$300 picosecond constant term is presented).
		ATLAS LAr test beam studies and an analysis of systematics indicate a goal of order 60-80 picosec could be achieved with significant effort\cite{Cleland}.

	The ATLAS ZDC, which has a projective Shashlik Quartz/Tungsten geometry, demonstrated  $\sigma_t$<100 picosec for high energy showers\cite{Otranto} and we are currently re-analyzing the data to understand limits to ATLAS ZDC time resolution.
	
	Other collider detectors have emphasized calorimeter time resolution. The PHENIX EM Calorimeter was designed to also perform PID using time of arrival and achieved $\sigma_t\sim$90 picoseconds for EM showers. The proposed ORKA detector at FNAL has a goal which is significantly ("several 10's of picoseconds") better\cite{Orka}.

	Phase II upgrade planning in the CMS Forward Calorimetry Group presents an opportunity to incorporate timing in new calorimetry which will likely be installed for HL-LHC. Although calorimeter concepts are still at an early stage in performance modeling, it appears likely that an upgraded forward calorimeter will emphasize jet measurement rather than dedicated high resolution EM measurement and capitalize on past investment (ILC motivated) in ultimate resolution jet calorimeters (with the ILC goal of resolving Z,W through hadronic decays).
	
	Currently there are 2 Dual Readout Calorimetry concepts being developed, both of which are likely to have a projective readout, as studied in RD52\cite{Wigmans}. Projective readout, in which particularly Cerenkov sampling Quartz fibers traverse the depth of the calorimeter and point back to the interaction region enable the same timing benefits that were observed in the ATLAS ZDC. As shown by RD52 (and the earlier SPACCAL project)\cite{Wigmans} the time structure of the light signal arriving at photosensors coupled to the front or back of the calorimeter allow good electron/hadron discrimination. Early studies also indicate $\sim30$ picosecond time of arrival spread leading photons from EM showers\cite{Nural} and it is possible that this will be useful for pileup mitigation.
	
	Given the importance of pileup mitigation going forward, we have proceeded also with R$\&$D on dedicated timing detectors, as discussed below. This would provide an alternative path should pileup mitigation through timing not be possible with the eventual choice of calorimeter-either for technical reasons or because of other factors which drive the calorimeter technology choice. It also enables work to proceed, necessary because the goal is an order of magnitude higher precision than has been achieved in ATLAS or CMS so far, in gaining experience with other tools (clock distribution, digitization and testing) which will be necessary to realize the fast timing goal.

\section{Tools}
	Literature on the general topic of optimal timing is mostly to be found outside the field of High Energy Physics. Aside from Information theory most of the technique is discussed in the literature of "Time Delay Estimation" with applications to radar or of delayed neuronal response to stimulus as a possible bio-marker for brain disease. In the sections below we focus on the aspects which are specific to the very short time scales specific to the LHC problem.
\subsection{Clock Distribution}
	At the required precision current clock distribution, based on fiber distribution of the LHC clock to the experiments and local beam pickups, needs to be enhanced because of thermal drifts in electronic components and in fiber ($\sim$20 picoseconds/$^oC$/km). 
	
	There is a lot of experience in clock distribution, based on interferometrically stabilized optical fibers (to a precision several orders magnitude better than we need) in the ILC and FEL community. At the same time this community has developed tools which would be sufficient for picosecond measurements--eg "the White Rabbit Project"\cite{Byrd}.
	Our group also designed a distribution system, capable of 5 picosecond stabilization, with a cost of $\sim\$60k$ for the FP420 R$\&$D project\cite{FP420}.
	
	So it is likely that the necessary tools are available and it is just a matter of choosing among the options.
\subsection{Digitization}
	Different approaches to digitization- waveform digitizers and precision TDC's with LHC architecture- are now achieving precision of $\sim3$ picoseconds at the ASIC level and expect to yield <10 picoseconds at the system level\cite{Herve}. So, again, it is likely that it is a matter of choosing among approaches.
\subsection{Testing}
	In addition to using conventional beam facilities at LNF, PSI and the CERN North Area, our group developed a dedicated facility for precision timing studies which capitalizes on the rms $\sigma_t\sim3$ picosecond bunch length of high  brightness electron Linacs (such as LAL or the BNL Accelerator Test Facility)\cite{singleelectron}. This is primarily useful for one-of-a-kind device testing.
	
	For detailed device testing, for example to map timing distortions due to the "weighting field" in Silicon devices, a femtosecond IR laser model for signal deposition by charged particles has proved invaluable. Since at a wavelength of $\sim980$ nm the absorption length in Si is $\sim1$mm, this provided ultimate time jitter and mapping at a spatial precision of $\sim20$ microns.
	
	Similarly, for a  gas Micromegas/transparent photocathode solution, a fast UV laser is being used to analyze the detector timing limits.
\section{Detectors}
	A 2006 paper based on Belle detector R$\&$D that demonstrated charged particle timing precision of $\sim5$ picoseconds was very influential. The Nagoya group obtained these results using Cerenkov light produced in the window (optimum thickness$\sim1$cm) of a MicroChannelPlate(MCP) PMT\cite{nagoya}.
	
	In principle this could be extended to a time-of-flight array composed of close-packed MCP's and this is a long-term goal of an R$\&$D effort in the US-LAPPD\cite{psec}. However commercially available MCPs are not currently suitable for a CMS or ATLAS application for several reasons. Aside from being costly, MCPs have a relatively short lifetime (both the photocathode and the MCP structure). While current research focuses on ameliorating this historical issue with MCPs, the lifetime requirements are orders of magnitude beyond what is now commercially available. Our group evaluated pre-production devices which are now commercially available from Hamamatsu (Hybrid Avalanche Photo Diodes-HAPDs) and demonstrated 11 picosecond single photon response\cite{Thomas}. This is similar to MCP performance and an order of magnitude better than the latest SiPMs available from Hamamatsu. Lifetime tests by Hamamatsu on their MCPs and on the HAPD device show that the latter has a lifetime roughly 3 orders of magnitude higher\cite{Hamamatsu} as required for an LHC application.
\subsection{Direct Charged Particle Detection}
	The excellent timing performance of the HAPD demonstrates that an APD itself is potentially capable of $\sim10$ picosecond charged particle timing. Internally, an HAPD accelerates a charged particle (photoelectron) which is then detected by an APD target with a resolution of <10 picoseconds.
	
	Before focusing on Si devices, we discuss general considerations for signal formation and collection in a gas or solid state charged particle sensors which affect timing jitter. 
\begin{itemize}
\item{The stochastic nature of signal formation due to Landau fluctuation in local energy deposition introduces a random structure in the signal which is formed at the electrodes. This results in a timing jitter relative to the actual time of the particle passage through the medium.}
\item{In addition, field distortions (i.e. the "weighting field" in a planar silicon detector or drift field in a gas chamber) can cause a \underline{systematic} shift in the signal arrival.}
\end{itemize}

	A large LHC detector- the ALICE TOF- overcomes the first effect in a gas chamber solution by making multiple measurements of the same track in an MRPC. ALICE TOF achieves $\sim80$ picosecond jitter at the system level but their current R$\&$D has achieved a time resolution of $\sim16$ picoseconds. Up to now, however, the ALICE technology has too low a rate capability for our application.
	
	An alternative approach, based on boosting the signal going into a Micromegas structure would, however, be capable of adequate rates\cite{Giomataris}.
	
	Aside from the Micromegas solution, the most promising technology for high rate picosecond timing seems to be Silicon devices, partly because of their versatility, as we will see below. CVD diamond trackers, for example, are limited by their intrinsic signal size, since no structures exist with internal gain and the best timing resolution for minimum ionizing particles (MIPs) is $\sim90$ picoseconds.
	
	Recently the NA62 experiment studied sources of timing jitter in a large Si tracker project (GTK). Their Si detectors consist of 200 micron thick planar detectors with no internal gain. A detailed study of test beam performance (where a timing jitter of 180 picosecond was obtained) showed that the dominant source of jitter, for their case, is non-uniformity in the weighting field. As a timing detector it would be nice to idealize a Silicon device as a detecting and, possibly, amplifying structure between 2 capacitor plates that have a)plenty of skin depth at a GHz. and b)maintain good transverse uniformity of the weighting field. This is, of course, rarely the case- particularly in a pixilated detector- and many of the recent improvements in SiPM timing performance, for example, come from improvements in metallization\cite{Hamamatsu}.

\begin{figure}[htb]
\begin{center}
 \includegraphics[width=0.49\textwidth]{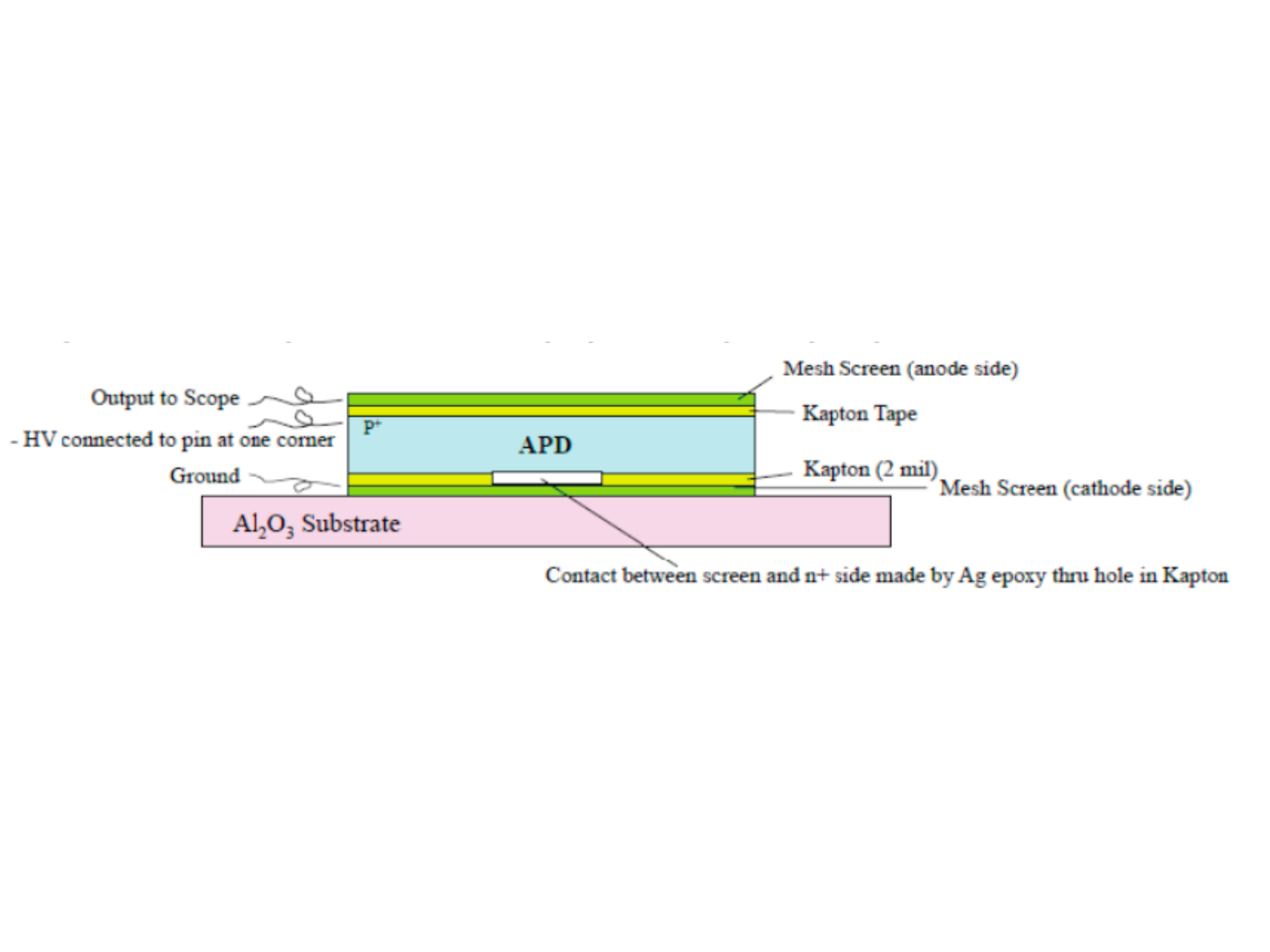}
  \includegraphics[width=0.49\textwidth]{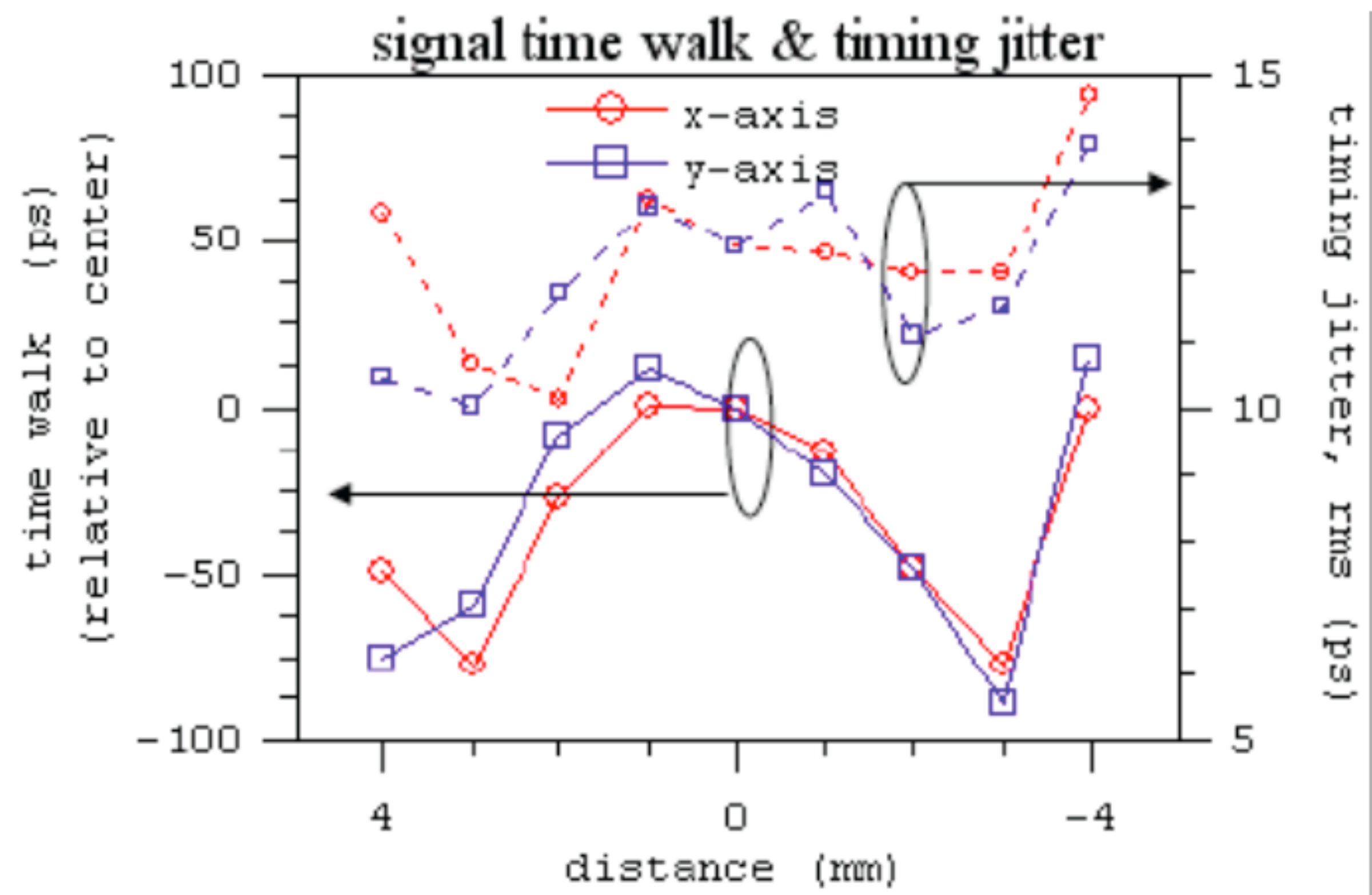}
  \caption{
     ({\it left}) Structure of the Deep-Depleted APD/MicroMegas design used for our measurements.
         ({\it right}) Timing performance with this design. The right hand vertical scale shows timing jitter as a 
         function of position along the (8 mm wide) detector centerline.
       }
  \label{fig2}
 \end{center}
\end{figure}

	In the case of our work with Deep depleted APDs, we went through a number of iterations, including Indium and Al vapor deposition, until we arrived at a solution with MicroMegas mesh acting as external capacitor plates (see Fig. 2a). As can be seen from Fig. 2b, this solution almost completely removes the variation with arrival time vs. impact position (initially >0.6 nsec). It also reduces the time jitter for a given impact position (studied using femtosecond IR laser).
	
	The laser model for MIP signal deposit, though useful for studying uniformity, misses a key aspect of charged particle interaction in Silicon. The very non-Gaussian, Landau/Vavilov distribution of signal deposit in thin Silicon also introduces a jitter as we will see from the following exercise.
	
	For our purposes, it is useful to think of the RMD deep-depleted structures as a 60 micron depletion region, of which only the "last" 40 microns has a saturated field, relevant for fast timing. Electron-hole pairs created by the traversing charged particle immediately start to drift towards the corresponding electrodes, inducing a signal, according to Ramo's Theorem. However, for our purposes and unlike the GTK case, what is significant is the electrons drifting (at $\sim10$ picoseconds/$\mu$m.) toward the amplification region where they are amplified by a factor of typically 500.
	
\begin{figure}[htb]
\begin{center}
 \includegraphics[width=0.49\textwidth]{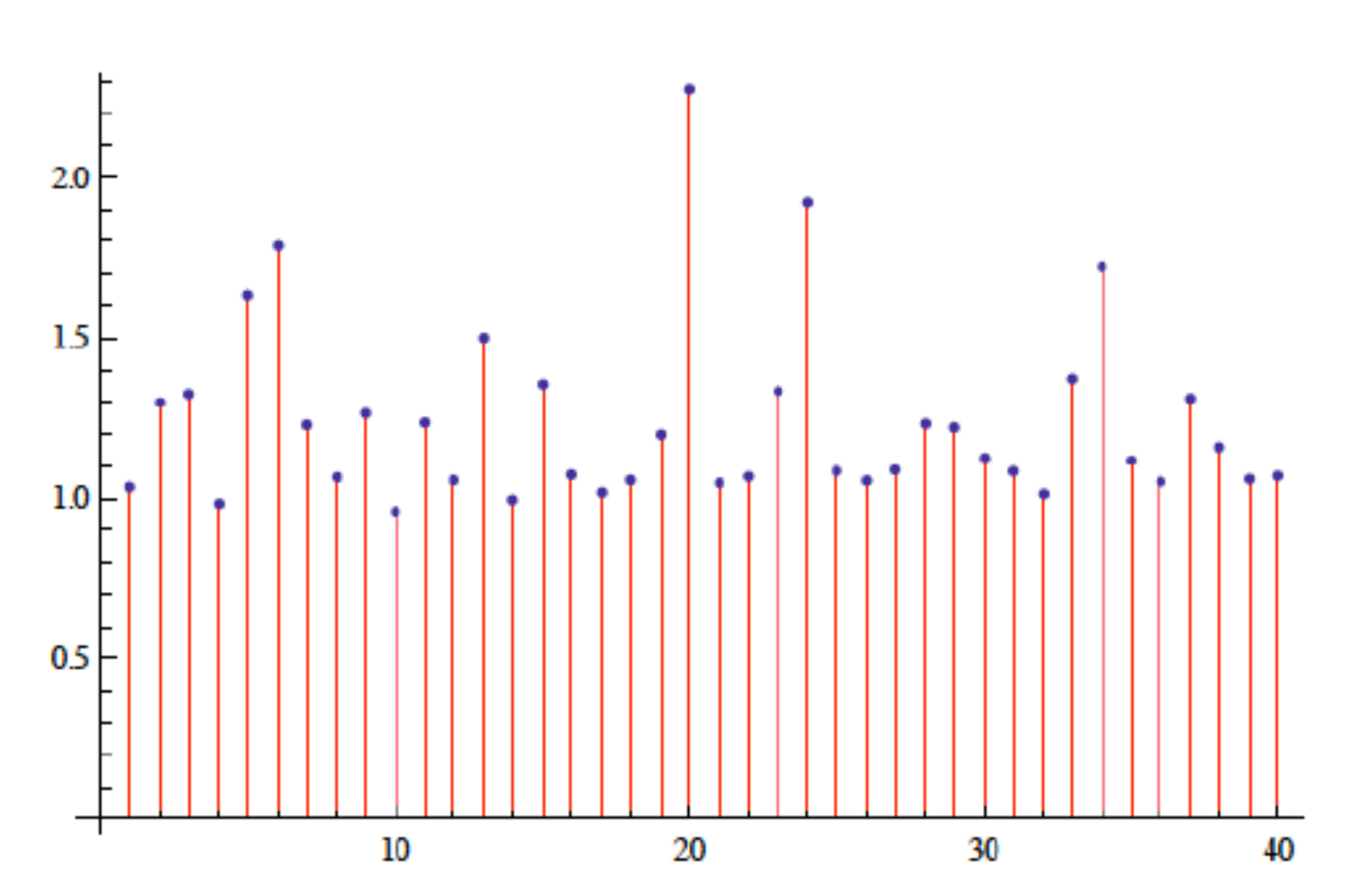}
  \includegraphics[width=0.49\textwidth]{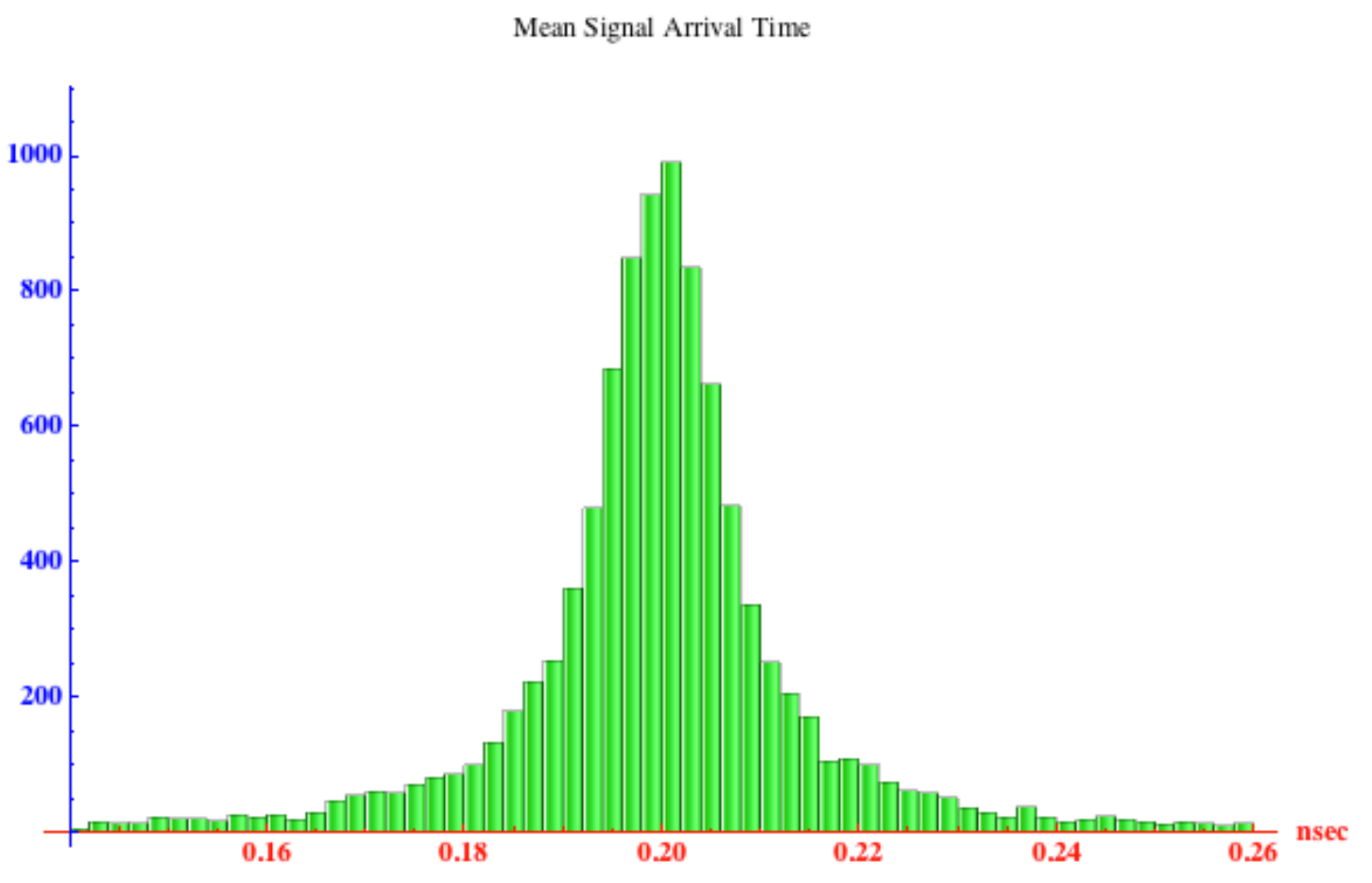}
  \caption{
     ({\it left}) Energy deposit (arbitrary scale) in sach of 40 1$\mu$m. layers.
         ({\it right}) Distribution in mean time of arrival relative to track time for 10,000 events.
       }
  \label{fig2}
 \end{center}
\end{figure}

	To model the jitter induced by Landau/Vavilov fluctuations, let's treat the 40$\mu$m region as 40- 1$\mu$m thick layers in each of which charge is deposited according to the correct Landau Distribution (see a typical event in Fig.3a). The non-uniform energy deposition results in time jitter of the energy-weighted time of arrival (and therefore a jitter in signal processing that
	is based on mean time of arrival) of about 21 picoseconds (see Fig. 3b). Constant fraction timing algorithms give essentially the same result.
	
	As expected, the jitter decreases in direct proportion to the depletion depth. For this reason, it might seem attractive to reduce the depth significantly\cite{Hartmut} but clearly such an optimization must take into account also the reduction in signal with reduced detector thickness. Our project works between the limits of modest gain ($\sim50$ and depletion depth of about 8 $\mu$m) and the high gain deep depletion devices discussed above, which we favor.
	
	Currently we are preparing for beam tests of these devices with significantly improved noise suppression and better matching of the pre-amplifier design to the larger ( 60 pF) capacitance of the detectors we now favor. Up to now, we have not reached the basic limits to time jitter from signal formation and noise due to detector leakage current. However Intermediate results can be found in the more detailed presentations on which this paper is based\cite{talks}.
\section{Next Steps}

	During the remainder of 2013 we expect to have completed beam measurements which will establish the timing resolution for MIPs with these detectors. In doing so we will also evaluate the latest electronics options which are now becoming available- including SAMPIC\cite{Herve}, DRS5 and possibly an early  HPTDC. We will also assess the timing algorithms which we have been using in beam tests up to now (but with lower SNR). Detector irradiation studies have been carried out at the CERN PS but the general issues regarding the lifetime of the APDs in the LHC environment have been discussed in an earlier paper\cite{design}.
	
	Concurrently, work will go forward assessing limits to calorimeter timing performance using existing data (particularly the ATLAS ZDC because of its relevance to the DRC options). This work falls within the wider context of general CMS physics simulation activities to determine the benefits of high precision timing for pileup mitigation and the extent to which the various calorimeter options lend themselves to timing measurement.

\acknowledgments
	I'd like to thank the organizers of the CHEF 2013 for a very interesting venue of "Paris in April" and also the Organizers of the Corfu Summer Institute. We are grateful for significant advice from Jan Kaplon, Pierre Jarron, Herve Grabas and Massimiliano Fiorini.


\begin{thebibliography}{10}
\bibitem{Huson} R.Huson, L.M.Lederman and R. Schwitters, "A Primer on Detectors in High Luminosity Environment",
in Snowmass 1982, Proceedings, Elementary Particle Physics and Future Facilities, pp. 361-368.
\bibitem{doeadrd} "Fast Timing Detectors for High Rate Environments" DOE Advanced Detector R$\&$D award, 2011.
\bibitem{correlation} S. White, "On the Correlation of Subevents in the ATLAS and CMS/Totem
Experiments", \begin{verbatim}http://arxiv.org/abs/0707.1500 Aug. 2007.\end{verbatim}
\bibitem{Thomas} T. Tsang and S. White, presented at "Making Single Molecule
Fluorescence (Lifetime) Measurements Simple" conference.
\bibitem{Fartoukh} S. Fartoukh, private communication.
\bibitem{Otranto} S. White, "Very Forward Calorimetry at the LHC- Recent Results from ATLAS" , Diffraction 2010, Proceedings, Otranto, Italy and
\begin{verbatim}http://library.wolfram.com/infocenter/Articles/7716/\end{verbatim}
\bibitem{Cleland} W. Cleland, private communication.
\bibitem{Orka} see: Intensity Frontier Instrumentation Report. \begin{verbatim}http://arxiv.org/pdf/1309.6704v1.pdf\end{verbatim}
\bibitem{Wigmans} \begin{verbatim}http://highenergy.phys.ttu.edu/dream/\end{verbatim}
\bibitem{Nural} Nural Akchurin and Chris Cowden, private communications.
\bibitem{Byrd} John Byrd, private communication. 
see also: \begin{verbatim}http://en.wikipedia.org/wiki/The_White_Rabbit_Project\end{verbatim}
\bibitem{FP420} \begin{verbatim}http://www.fp420.com/papers/fp420.pdf\end{verbatim}
\bibitem{singleelectron} BNL ATF Experiment AE-55, S. White and K. McDonald, co-spokespersons.
\bibitem{nagoya} Kenji Inami et al. ,ÒA 5 psec TOF-counter with an MCPÐPMTÓ, Nuclear Instruments and Methods A 560 2 (2006) 303-308.
\bibitem{psec} Large Area Picosecond Photo-Detectors Project. psec.uchicago.edu
\bibitem{Hamamatsu} Dave Fatlowitz and Ardavan Ghassemi, private communication. Most of the reference to performance limits of commercial MCP, SiPM and
HAPD devices in the current paper follows from a review we did with Hamamatsu in preparation for the Jan. 2013 CPAD meeting. An attempt was made to do
a realistic evaluation of lifetime etc. based on 1 manufacturers experience with all of these technologies.
\bibitem{Giomataris} E. deLagnes, I. Giomataris, R. Veenhof and S. White- paper in preparation.
\bibitem{Hartmut} Hartmut Sadrozinski, private communication. See also his 2013 Snowmass White paper and DPF presentation.
\bibitem{talks} S. White presentations at Chef2013 and Corfu Summer Institute 2013;
\begin{verbatim}www.physics.ntua.gr/corfu2013/lectures.html\end{verbatim}
\bibitem{Herve} H. Grabas presentation at CERN August 27, 2013
\begin{verbatim}http://indico.cern.ch/getFile.py/access?contribId=19&sessionId=4&resId=0&materialId=slides&confId=267313\end{verbatim}
and private communications also w. Stefan Ritt, Eric deLagnes and Jorgen Christiansen.
\bibitem{design} White,~S. et al "Design of a 10 Picosecond Time of Flight Detector using Avalanche PhotoDiodes", Jan. 2009, \begin{verbatim}http://arxiv.org/abs/0901.2530.\end{verbatim}

\end{thebibliography}
\end{document}